\documentstyle[12pt,fullpage,subeqn]{article}
\begin{document}
\thispagestyle{empty}
\catcode`@=11 \@addtoreset{equation}{section} \catcode `@=12
\renewcommand{\theequation}{\thesection.\arabic{equation}}
\def\lsim{\:\raisebox{-0.5ex}{$\stackrel{\textstyle<}{\sim}$}\:}
\def\gsim{\:\raisebox{-0.5ex}{$\stackrel{\textstyle>}{\sim}$}\:}
\def\lapprox{{\underrel{\scriptstyle<}\over\sim}}
\def\lessapprox{{\buildrel{<}\over{\scriptstyle\sim}}}
\def\be{\begin{equation}}
\def\ee{\end{equation}}
\def\beq{\begin{eqnarray}}
\def\eeq{\end{eqnarray}}
\def\eq#1{(\ref{#1})}

\def\np#1#2#3{Nucl. Phys. {\bf B#1} (#2) #3} 
\def\pl#1#2#3{Phys. Lett. {\bf #1B} (#2) #3}
\def\prl#1#2#3{Phys. Rev. Lett. {\bf #1} (#2) #3}
\def\physrev#1#2#3{Phys. Rev. {\bf D#1} (#2) #3}
\def\prd#1#2#3{Phys. Rev. {\bf D#1} (#2) #3}
\def\ap#1#2#3{Ann. Phys. {\bf #1} (#2) #3}
\def\prep#1#2#3{Phys. Rep. {\bf #1} (#2) #3}
\def\rmp#1#2#3{Rev. Mod. Phys. {\bf #1} (#2) #3}

\pretolerance=1000000

\begin{flushright}
TIFR/TH/99-54\\
CERN-TH/99-321\\
October 1999\\
hep-th/9910194
\end{flushright}
\begin{center}
\vspace{2 ex}
{\large{\bf D1/D5 System with B-field, Noncommutative Geometry and
the CFT of the Higgs Branch}}\\
\vspace{3 ex}
Avinash Dhar$^a$, Gautam Mandal$^{a,c}$,
Spenta R. Wadia$^{a \dagger}$\\
and K.P. Yogendran$^a$\\
~ \\
$^a${\sl Department of Theoretical Physics,
Tata Institute of Fundamental Research,}\\
{\sl Homi Bhabha Road, Mumbai 400 005, INDIA.} \\
~ \\
$^c$ {\sl CERN Theory Division, CH-1211, Geneva 23, Switzerland.}
\\
\vspace{10 ex}
\pretolerance=1000000
\bf ABSTRACT\\
\end{center}
\vspace{1 ex}

The D1/D5 system is considered in the presence of the NS B field. An
explicit supergravity solution in the asymptotically flat and near
horizon limits is presented. Explicit mass formulae are presented in
both cases. This solution has no D3 source branes and represents a
true bound state of the D1/D5 system. We study the motion of a
separated D1-brane in the background geometry described above and
reproduce the Liouville potential that binds the D1 brane. A gauge
theory analysis is also presented in the presence of Fayet-Iliopoulos
(FI) parameters which can be identified with the self-dual part of the
NS B field. In the case of a single D5-brane and an arbitrary number
of D1 branes we can demonstrate the existence of a bound state in the
Higgs branch. We also point out the connection of the SCFT on the
resolved Sym$_{Q_1Q_5}(\tilde T^4)$ with recent developments in
non-commutative Yang-Mills theory.

\vfill
\vspace{5ex}
\hrule
\begin{flushleft}
\baselineskip 2ex
{\small e-mail:  adhar,mandal,wadia,kpy@theory.tifr.res.in \\
$^\dagger$ Jawaharlal Nehru Centre for Advanced
Scientific Research. Bangalore 560012, INDIA.}
\end{flushleft}
\baselineskip 3.9ex

\clearpage

\section{Introduction}

The D1/D5 system of type IIB string theory compactified on $S^1 \times
T^4$ (or type IIA on $S^1 \times K^3$) has been the subject of many
investigations since it was introduced by Strominger and Vafa
\cite{VafStr96} in their work on the microscopic origin of the entropy
of the corresponding black hole solution in (4+1)-dimensions.
This system is amenable to a rather precise analysis because its
dynamics in the limit $\alpha' \rightarrow 0$ is described by a
non-abelian gauge theory. This fact underlies the hope that a
correctly understood and calculable theory of Hawking radiation can
emerge for the near extremal 5-dim. black hole. This hope is
realized for the emissivity of the massless modes upto an over all
numerical constant which can be fixed by assuming the AdS/CFT duality
\cite{DavManWad98}.

It turns out that the infra-red dynamics of D1/D5 system is
described by a 2-dimensional superconformal sigma model whose target
space is the moduli space of Yang-Mills instantons on $T^4$ or $K^3$.
In this paper we will will restrict ourselves to $T^4$. This SCFT has a
singularity
corresponding to the vanishing size of instantons which corresponds to
the origin of the Higgs branch. The singularity is resolved by
turning on any one of the four exactly marginal operators
which correspond to the blow-up modes. In the SCFT
with target space Sym$_{Q_1Q_5}(\tilde T^4)$
\footnote{%
  Sym$_k M$ denotes the symmetric product $ M^k/S(k)$ where $S(k)$ is
  the permutation group of $k$ elements. The tilde on $T^4$ is a
  reminder that the 4-torus is not necessarily identical to the
  original $T^4$ \cite{DavManWad98}.},
these 4 operators arise from
the twisted sector corresponding to
a single non-trivial cycle of length 2 and are denoted by $\tau_0$
and $\tau_1$ in \cite{DavManWad99}. Using a global symmetry
of the SCFT these were identified
in \cite{DavManWad99} to be dual
to the four  supergravity fields: $B^{(+)}$
(self-dual part of the NS 2-form $B$) and $aC^{(0)} +
bC^{(4)}$ (a linear combination of the Ramond-Ramond
forms $C^{(0)}$ and
$C^{(4)}$). Seiberg and Witten \cite{SeiWit99a} have shown
that the singularity of the Higgs branch corresponds to the decay of
the marginally stable bound system of D1/D5 branes to sub-systems of
D1/D5 branes. Turning on any 4 marginal operators evades the
singularity of the Higgs branch, because the marginally stable bound
state is now expected to become a true bound state.

This fact can be argued in various ways and if we assume the AdS/CFT
conjecture it certainly implies the existence of a black hole solution
in the bulk where the moduli fields corresponding to the blowing up
modes are non-zero. Such a black hole will be stable against
fragmentation into constituents. In this paper we present such a
solution for which $B \neq 0$ (in particular, $B^{(+)}|_{\rm brane}
\neq 0$). A preliminary version was already reported in
\cite{Spain-talk}. We present a detailed analysis of the
charges, both at infinity and at the horizon, to ensure that the
system is indeed D1/D5 and there are no source D3-branes.
Given this new solution we repeat the analysis in \cite{SeiWit99a} and
study the motion of a test D1 brane in the background of this
solution. We demonstrate the Liouville potential that binds the D1
brane to a large number of D1/D5 branes. The coefficient of the
Liouville potential is proportional to $b^{2}$, where $b$ is the
self-dual part of the NS B field at the horizon.

The discussion of the bound state in the context of the microscopic
theory can be done efficiently by approximating the instanton sector
of the 5-brane gauge theory by the ${\cal N}=4$, $U(Q_1)\times U(Q_5)$
gauge theory in 2 dims \cite{CalMal, HasWad97}. We analyse the
D-flatness (ADHM) equations when the Fayet-Iliopoulos (FI) parameters
are non-zero, and indicate that the $(1,5)$ and $(5,1)$ strings
condense on the hyper-Kahler manifold $T^{*}CP^{N-1}$, where
$N=Q_1Q_5$. We then discuss the transverse fluctuations of the brane
system (case $Q_5=1$) and show that these do not have flat directions
transverse to the brane system.  If the FI parameters are
non-zero. This demonstrates a normalizable ground state of the Higgs
branch.

As this work was getting completed we became aware of the recent
developments in non-commutative geometry and string theory:
\cite{SeiWit-talk,SeiWit99b,Connes,DougHull,BigSuss,Schom,Krogh,Ardalan,
HashItz,Mal_talk,MalRuss,ChuHo}. The presence of the $B$ in the world volume
of a collection of D-branes changes the boundary conditions of the
open strings that mediate the interactions of the branes. The effect
of this is that the Yang-Mills (YM) theory in 6-dimensions in $T^4
\times S^1 \times R^1$ that describes the collective modes of the
D1/D5 system is now replaced by a noncommutative YM theory (NCYM),
with the noncommutativity parameter determined by $B$, as explained in
\cite{SeiWit99b}. The moduli space to consider should now be that of
the self-dual gauge fields in NCYM on the four-torus: ${\hat F}^{(-)}
= 0$, where $\hat F$ denotes the NCYM field strength and the
superscript $^{(-)}$ denotes anti-self-dual part, taken with respect
to an appropriate (open string) metric. These conditions once again
preserve 8 SUSY's and can be considered as a deformation of the
original self dual equations.

In case of the YM theory in $R^4$, Nekrasov and Schwarz \cite{NekSch}
have shown, by an analysis of the ADHM equations, that the moduli
space of ${\hat F}^{(-)} = 0$ is equivalent to a resolution of the
moduli space of the ordinary $F^{(-)} = 0$ by the $B^{(+)}_{ij}$
(Fayet-Iliopoulos) parameters.
For the case of the torus $T^4$ or $K^3$ we are unaware of a similar
precise statement but an educated guess is the following:
The moduli space of ${\hat F}^{(-)}_{ij} = 0$ on the torus is a
resolution of the symmetric product: Sym$_{Q_1Q_5}(\tilde T^4)$. In
applications, for instance, when one constructs the orbifold CFT on
the symmetric product one implicitly assumes that a certain flux
persists in the squashed 2 cycles. This flux corresponds to turning on
the
operator $\tau_0$ in \cite{DavManWad99}. We conjecture that the SCFT
obtained by further adding the 3 operators $\tau_1$, with arbitrary
values of
the corresponding moduli, is the SCFT where target space is the moduli
space of $F^{(-)}_{ij} = 0$.

The paper is organized as follows: We first construct in Section 2, the
supergravity solution that corresponds to the D1/D5 system in the
presence of non-zero $B$ and the Liouville potential
felt by a probe D1 brane. In Section 3, we discuss the ${\cal N}=4$,
$U(Q_1)\times U(Q_5)$ gauge theory in 2 dimensions when the
Fayet-Iliopoulos (FI) parameters are non-zero and
discuss the transverse fluctuations of the brane system (case $Q_5=1$).
We conclude with a proposal for the sigma model on the
moduli space of NCYM instantons.
\vspace{1em}

\section{D1/D5 System with B-field}

In this section we will construct a solution
\cite{Spain-talk} to type IIB supergravity
compactified on $T^4$ which represents the D1/D5 system with a NS
B-field in the internal $T^4$ directions. We will obtain this solution
from brane systems without a B-field by a series of T-dualities and use
it to study  various properties of this system discussed in the
Introduction.

We will denote the $T^4$ directions as $x^{6,7,8,9}$.  We begin with a
IIB supergravity solution corresponding to two sets of orthogonal
D3-branes extending along $x^{5,6,7}$ and $x^{5,8,9}$ respectively.
The common direction, $x^5$, is assumed non-compact for the moment.
The first set of D3-branes, which wraps the $T^4$ in $x^6$ and $x^7$
directions, is ``smeared'' (see, e.g. \cite{Gauntlett-review})
 over the remaining $T^4$ directions, $x^8$
and $x^9$. The other set of D3-branes, which wraps $x^8$ and $x^9$
directions, is smeared over $x^6$ and $x^7$. This solution
\cite{Tseytlin-harmonic-superposition} is
given by the following metric, dilaton and self-dual 5-form RR field
strength:
\beq
\label{1}
ds^2 &=& (f_1f_5)^{-1/2} (-dt^2 + (dx^5)^2) + (f_1f_5)^{1/2} (dr^2 +
r^2d \Omega^2_3) \nonumber \\
&& + \left({f_1\over f_5}\right)^{1/2} (dx^6)^2 + (dx^7)^2) +
\left({f_5\over f_1}\right)^{1/2} ((dx^8)^2 + (dx^9)^2) , \nonumber \\
e^{2\phi} &=& 1, \nonumber \\
F^{(5)} &=& K^{(3)} \wedge dx^6 \wedge dx^7 + \tilde K^{(3)} \wedge
dx^8 \wedge dx^9, \nonumber
\eeq
where
\beq
\label{2}
K^{(3)} &=&  (-df^{-1}_5 \wedge dt \wedge dx^5 + \alpha' \mu_1
\epsilon_3) , \nonumber \\
\tilde K^{(3)} &=&  (-df^{-1}_1 \wedge dt \wedge dx^5 + \alpha' \mu_5
\epsilon_3) . \nonumber
\eeq
Here $\epsilon_3$ is the volume form of the 3-sphere and the harmonic
functions $f_1$ and $f_5$ are given by
\be
\label{3}
f_{1,5} = 1 + {\mu_{1,5} \over 2} \left({\alpha' \over r^2}\right) ,
\ee
where $\mu_{1,5}$ are dimensionless numbers related to the numbers of
D3-branes in the two sets. We have used here a notation which is
appropriate to the D1/D5 solution with a B-field that we are
ultimately interested in. For the sake of completeness, we also give
here the type IIB low-energy action to which the above is a solution:

\beq
\label{action}
S_{IIB} &=& {1\over 2\kappa^2} \int d^{10} x \sqrt{-G} \bigg\{
e^{-2\phi}
\left(R + 4 (\nabla \phi)^2 - {1\over 2.3 !} (H^{(3)})^2\right) -
{1\over 2.3 !} (F^{(3)})^2 \nonumber \\ [2mm]
&& - {1\over 4.5 !} (F^{(5)})^2\bigg\} + {1\over 4\kappa^2} \int C^{(4)}
\wedge F^{(3)} \wedge H^{(3)} ,
\eeq
where $(H^{(3)})^2 = H^{(3)}_{\rm MNP} H^{(3){\rm MNP}} , \ (F^{(n)})^2
=
F^{(n)}_{M_1 \cdots M_n} F^{(n)M_1 \cdots M_n}$ and, using the standard
form notation,
\be
\label{4}
H^{(3)} = dB^{(2)}_{NS} , \ \ F^{(3)} = dC^{(2)} , \ \ F^{(5)} =
dC^{(4)} - {1\over 2} C^{(2)} \wedge H^{(3)} + {1\over 2} B^{(2)} \wedge
F^{(3)}.
\ee
The self-duality constraint, $\ast F^{(5)} = F^{(5)}$, is imposed at
the level of the equations of motion. Also, $\kappa^2 = 8 \pi G_{10}$,
where
$G_{10} = 8\pi^6 g^2 \alpha^{\prime 4}$ is the 10-dimensional Newton's
constant (in the convention that the dilaton, $\phi$, vanishes
asymptotically).

The desired D1/D5 solution is obtained from the above solution by the
following series of T-duality transformations: A rotation in the $x^6$
- $x^8$ plane by an angle $\varphi$, followed by a T-duality in the
(new) direction $x^8$, followed by a rotation in the $x^7$ - $x^9$ plane
by an
angle $\psi$, which is finally followed by a T-duality in the (new)
direction $x^9$. We have used the T-duality rules for RR field
strength in the form given in \cite{Fawad}, adapted to our
conventions (self-dual 5-form RR field strength). We present below the
solution obtained in this way:
\beq
\label{full-solution}
ds^2 &=& (f_1f_5)^{-1/2} (-dt^2 + (dx^5)^2) + (f_1f_5)^{1/2} (dr^2 +
r^2 d \Omega^2_3) \nonumber \\ [2mm]
&& + (f_1f_5)^{1/2} \left\{ Z^{-1}_\varphi ( (dx^6)^2 + (dx^8)^2) +
Z^{-1}_\psi ((dx^7)^2 + (dx^9)^2) \right\} , \\ [2mm]
\label{full-solution-1}
e^{2\phi} &=& f_1 f_5 / Z_\varphi Z_\psi , \\ [2mm]
\label{full-solution-2}
B^{(2)}_{NS} &=& (Z^{-1}_\varphi \sin\varphi \cos\varphi (f_1-f_5) +
b_{68})
dx^6 \wedge dx^8 \nonumber \\ [2mm]
&& + (Z^{-1}_\psi \sin\psi \cos\psi (f_1-f_5) + b_{79}) dx^7 \wedge
dx^9 , \\ [2mm]
\label{full-solution-3}
F^{(3)} &=& \cos\varphi \cos\psi \tilde K^{(3)} + \sin\varphi \sin\psi
K^{(3)} , \\ [2mm]
\label{full-solution-4}
F^{(5)} &=& Z^{-1}_\varphi (-f_5 \cos\varphi \sin\psi K^{(3)} + f_1
\cos\psi
\sin\varphi \tilde K^{(3)}) \wedge dx^6 \wedge dx^8 \nonumber \\ [2mm]
&&+ Z^{-1}_\psi (-f_5 \cos\psi \sin\varphi K^{(3)} + f_1 \cos\varphi
\sin\psi \tilde K^{(3)}) \wedge dx^7 \wedge dx^9 , \\ [2mm]
\label{full-solution-5}
Z_{\varphi, \psi} &=& 1 + {\mu_{\varphi,\psi} \over 2} \left({\alpha'
\over r^2}\right) , \ \ \ \mu_\varphi = \mu_1 \sin^2\varphi + \mu_5
\cos^2 \varphi , \ \ \ \mu_\psi =
\mu_1 \sin^2\psi + \mu_5
\cos^2 \psi.
\eeq
Here $b_{68}$ and $b_{79}$ are arbitrary constants which we have added
at the end by a T-duality transformation that
shifts the NS B-field by a constant. Note that for $\varphi = \psi =
0$ and $b_{68} = b_{79} = 0$, the above solution reduces to the
well-known solution for D1/D5 system without B-field.

\vspace{2em}

\subsection{Charges}

In this subsection we will discuss the various charge densities
\footnote{These are densities rather
than the total charges, as explained  below \eq{q3p}.}
associated with the RR fields.
The electric charge densities are given
by
\[
Q_{\rm elec} = {1\over 2\kappa^2} \oint \ast F^{(n)} ,
\]
and the magnetic charge densities by
\[
Q_{\rm mag} = {1\over 2\kappa^2} \oint F^{(n)} .
\]
The integrals above are evaluated at asymptotic distances in the
4-dimensional noncompact space. Note that for arbitrary $\varphi$ and
$\psi$ values, $F^{(3)}$ and $F^{(5)}$ have non-trivial
$r$-dependence which is
different from the asymptotic dependence $1/r^3$.  This means, in
particular, that the numerical values of the charges computed in the
asymptotically flat geometry are different from those computed in the
near horizon AdS geometry. We shall give below expressions for charges
in both cases. The charges in the near horizon geometry have been
distinguished from those in the asymptotically flat geometry by a
superscript $^{(h)}$.

The RR field strength $F^{(3)}$ gives rise to D1-brane and D5-brane
charges while $F^{(5)}$ gives rise to two types of D3-brane electric
and magnetic charges. For our solution these charge densities are, in
an obvious notation,
\beq
\label{q1}
Q_1 &=& \beta (\mu_1 \cos\varphi \cos\psi + \mu_5 \sin\varphi
\sin\psi) , \\ [2mm]
\label{q5}
Q_5 &=& \beta (\mu_5 \cos\varphi \cos\psi + \mu_1 \sin\varphi
\sin\psi) , \\ [2mm]
\label{q3}
Q_3 \equiv Q^{568}_3 &=& \beta (\mu_5 \cos\varphi \sin\psi - \mu_1
\cos\psi
\sin\varphi) , \\ [2mm]
\label{q3p}
Q'_3 \equiv Q^{579}_3 &=& \beta (\mu_5 \cos\psi \sin\varphi - \mu_1
\cos\varphi \sin\psi) ,
\eeq
where $\beta = {\alpha' \Omega_3}/ (2\kappa^2)$ and
$\Omega_3 = 2\pi^2$ is
the volume of a unit 3-sphere. Here $Q_1$ is the charge `density' of
D1-branes obtained by dividing in addition by the asymptotic volume of
the torus $T^4$
and $Q_3, Q'_3$ are D3-brane charge densities obtained by dividing in
addition by appropriate $T^2$ volumes.

In the near horizon geometry, the corresponding expressions for the
charge densities are,
\beq
\label{q1h}
Q^{(h)}_1 &=& \beta {\mu_1 \mu_5 \over \mu_\varphi \mu_\psi}
(\mu_5 \cos\varphi \cos\psi + \mu_1 \sin\varphi \sin\psi) , \\ [2mm]
\label{q5h}
Q^{(h)}_5 &=& Q_5 , \\ [2mm]
\label{q3h}
Q^{(h)}_3 \equiv Q_3^{568(h)} &=& \beta {\mu_1\mu_5\over \mu_\psi} \sin
(\psi - \varphi) , \\ [2mm]
\label{q3ph}
Q^{(h)'}_3 \equiv Q^{579(h)}_3 &=& \beta {\mu_1\mu_5 \over
\mu_\varphi} \sin (\varphi - \psi) .
\eeq
As expected, only the D5-brane charge density is the same in the two
cases.

\subsection{Source Branes and Mass}

We now require that our solution should correspond to only D1- and
D5-brane sources. That is, we want the D3-brane charges to be induced
purely by our nonzero B-field, and not by any source
D3-branes\footnote{Note that the full space of supergravity solutions
  generated by the T-duality group $O(4,4)$ will have arbitrary number
  of source D1-,D3- and D5-branes. We will choose below
  parameters of our T-duality transformations in such a way that we
  have  a `pure D1/D5 system'.}.

There are two contributions to the D3-brane charge induced by the
B-field. One comes from the bulk Chern-Simons term in (2.2),
\be
\label{5}
{1\over 4\kappa^2} \int C^{(4)} \wedge F^{(3)} \wedge H ,
\ee
and the other from the coupling
\be
\label{6}
Q_5 \int C^{(4)} \wedge B
\ee
in the Chern-Simons part of the D5-brane would volume action,
\be
\label{7}
\rho_{(5)} {\rm Tr} \int \sum_n e^{2\pi\alpha' F+B} \wedge C^{(n)} .
\ee
Here $\rho_{(5)}$ is the charge density of a D5-brane and the trace is
over the non-abelian gauge group ${\cal U}(n_5)$, $n_5$ being the
number of D5-branes.

The coupling in \eq{5} contributes a $\delta$-function term (with
support at the location of the D5-brane in the 4 noncompact
dimensions) in the local D3-brane charge density, proportional to the
value of the B-field at the horizon. On the other hand, the
contribution of \eq{6} to the D3-brane charge is proportional to the
difference of the asymptotic and the horizon values of B. It can be
shown that the sum of these two contributions is an induced D3-brane
charge proportional to the asymptotic value of the B field, thus
leading to \eq{induced-d3} below.

Similarly to the above, there are also contributions to the total
D1-brane charge which are induced by the B-field. Apart from the bulk
Chern-Simons term \eq{6}, there is also a contribution from the D5-brane

Chern-Simons action which is given by
\be
\label{8}
Q_5 \int {1\over 2 !} B \wedge B \wedge C^{(2)} .
\ee
In this case also the two contributions add up to finally give induced
D1-brane charge which depends only on the asymptotic value of B,
leading to \eq{induced-d1} below.

Let us now discuss the two cases of the asymptotically flat and the
near horizon geometry separately. In other words, we will consider
first the case of the full supergravity solution \eq{full-solution}
and then the case in which we take the near-horizon limit in the
spirit of \cite{Maldacena} and treat it as an independent solution of
IIB
supergravity in its own right.

\vspace{5ex}
\noindent(i) \underbar{Asymptotically flat geometry}
\vspace{5ex}

Here, the absence of any D3-brane sources is ensured by demanding that
\be
\label{induced-d3}
Q_3 = B^{(\infty)}_{79} Q_5 , \ \ \ Q'_3 = B^{(\infty)}_{68} Q_5 ,
\ee
where
\be
\label{9}
B^{(\infty)}_{79} = b_{79} , \ \ \ B^{(\infty)}_{68} = b_{68} ,
\ee
are the asymptotic values of the two nonzero components of the
B-field. Eqns. \eq{induced-d3} are satisfied if we set
\beq
\label{no-d3-1}
b_{68} &=& {\mu_5 \sin\varphi \cos\psi - \mu_1 \cos\varphi \sin\psi
\over \mu_5 \cos\varphi \cos\psi + \mu_1 \sin\varphi \sin\psi} , \\
[2mm]
\label{no-d3-2}
b_{79} &=& {\mu_5 \cos\varphi \sin\psi - \mu_1 \sin\varphi \cos\psi
\over \mu_5 \cos\varphi \cos\psi + \mu_1 \sin\varphi \sin\psi} .
\eeq
Furthermore, if $Q_{1s}$ denotes the D1-brane charge which arises from
source D1-branes, then, using \eq{action} and \eq{8} it can be shown
that
\be
\label{induced-d1}
Q_{1s} = Q_1 - B^{(\infty)}_{68} \ B^{(\infty)}_{79} \ Q_5 .
\ee
{}From (2.10)-(2.13) and (2.22)-(2.26) we then get
\be
\label{q1s}
Q_{1s} = \beta^2 {\mu_1 \mu_5 \over Q_5} .
\ee

\bigskip

\noindent\underbar{Mass}

The supergravity solution that we have is ${1\over 4}$ BPS. The mass of
this
${1\over 4}$ BPS solution
\cite{Obers-Lee}, which coincides with its ADM mass, is given
in terms of the appropriate charge densities by
\be
\label{10}
M^2 = (Q_1 + Q_5)^2 + (Q_3 - Q'_3)^2
\ee
Note that, as expected, the mass given by \eq{10} equals the ADM mass of

the intersecting D3-brane solution from which the present solution was
obtained by T-duality, i.e.
\be
M = \beta (\mu_1 + \mu_5) .
\ee

We wish to study $M$ as a function of the moduli, keeping $Q_{1s}$
and $Q_5$ fixed. This last requirement, together with
\eq{no-d3-1} and \eq{no-d3-2},
determines $\mu_1 , \mu_5 , \varphi$ and $\psi$ as functions of the
two moduli $b_{68}$ and $b_{79}$:
\beq
\label{11a}
\tan (\varphi \pm \psi) &=& {b_{68} \pm b_{79} \over 1 \mp b_{68}
b_{79} \mp Q_{1s}/Q_5} , \\ [2mm]
\label{11b}
\beta\mu_1 &=& {Q_5 \over 2} \bigg\{(1+ b_{68} b_{79} +Q_{1s}/Q_5) \sec
(\varphi -
\psi) \nonumber \\ [2mm]
&& \hspace{2cm} - (1- b_{68} b_{79}  - Q_{1s}/Q_5) \sec (\varphi +
\psi)\bigg\} ,
\\ [2mm]
\label{11c}
\beta\mu_5 &=& {Q_5 \over 2} \bigg\{(1+bb'+Q_{1s}/Q_5) \sec (\varphi -
\psi) \nonumber \\ [2mm]
&& \hspace{2cm} + (1- b_{68} b_{79} - Q_{1s}/Q_5) \sec (\varphi +
\psi)\bigg\} .
\eeq
For arbitrary $b_{68}$ and $b_{79}$, $\mu_1$ and $\mu_5$ given by the
above
equations will not satisfy any quantization conditions, unlike, e.g.,
for $b_{68} = b_{79} = 0$ for which $\mu_5 = 2g_sn_5$ and $\mu_1 =
\displaystyle{{2g_sn_1 \over V_{T^4}/(2\pi)^4 \alpha'^2}}$, where $n_1$
and $n_5$ are integers. This should not
unduly worry us since for the system under discussion only $Q_{1s}$
and $Q_5$ have a direct physical interpretation in terms of brane
sources, and it is only on
these that appropriate quantization conditions must be imposed.

We now wish to extremize \eq{10} with respect to both $b_{68}$ and
$b_{79}$. This requires
\be
\label{12}
b_{68} = - b_{79} = \pm \sqrt{Q_{1s}/Q_5 -1}  ,
\ee
and then the mass at the fixed point is given by
\be
\label{13}
M^2 = 4 Q_{1s} Q_5 .
\ee
For fixed asymptotic volume of $T^4$, Eqns. \eq{12} can be seen to fix
the $B$-moduli at the minimum in terms of the ratio of D1- and D5-brane
source charges. For a different value for this ratio, the mass is
minimized for appropriately different values of the $B$-moduli.
Stated differently, for a fixed value of the $B$-moduli,
unless the charge-ratio is given by \eq{12}, the system will
be unstable and decay to a system which satisfies \eq{12}.

Hence,  once the B-moduli are fixed at the values in \eq{12},
{\em the system forms a true bound state of the constituent branes}
\footnote{It is interesting to consider the case when the fragmentation
happens
  keeping the ratio of $Q_{1s}/Q_5$ fixed. Eqn. \eq{13} would predict
  zero binding energy and fragmentation at no cost. However, such an
  eventuality will be prevented for mutually prime $Q_{1s}, Q_5$.
  This is in accord with \cite{Vaf95gas,SeiWit99a,Dijkgraaf} which
indicate
  singularities associated with points in the moduli space where
  $Q{1s}$ and $Q_5$ have common factors.}.

\vspace{5ex}

\noindent (ii) \underbar{Near horizon geometry}
\vspace{5ex}

In this case, absence of D3-brane sources is ensured if we set
\be
\label{no-d3-nh}
Q^{(h)}_3 = B^{(h)}_{79} Q_5 , \ \ \ Q^{(h)'}_3 = B^{(h)}_{68} Q_5 ,
\ee
where
\beq
\label{14a}
B^{(h)}_{68} &=& {\mu_1 - \mu_5 \over \mu_\varphi} \sin\varphi
\cos\varphi + b_{68} , \\ [2mm]
\label{14b}
B^{(h)}_{79} &=& {\mu_1 - \mu_5 \over \mu_\psi} \sin\psi
\cos\psi + b_{79} ,
\eeq
are the horizon values of the two nonzero components of the
B-field. Eqns. \eq{no-d3-nh}
are satisfied for precisely the same values of
$b_{68}$ and $b_{79}$ as in \eq{no-d3-1} and \eq{no-d3-2}.
Moreover, we see that
in this case
\be
\label{15}
{B^{(h)}_{68}\over \mu_\psi} = - {B^{(h)}_{79} \over \mu_\varphi} ,
\ee
which is the self-duality condition on the B-field in the near horizon
geometry. We also note that the volume of $T^4$ at the horizon is given
by
\be
\label{16}
V^{(h)}_{T^4} = {\mu_1\mu_5 \over \mu_\varphi \mu_\psi} = {Q^{(h)}_1
\over Q_5} .
\ee

The D1-brane charge that arises from
source D1-branes in this case is given by
\be
\label{17}
Q^{(h)}_{1s} = Q^{(h)}_1 - B^{(h)}_{68} B^{(h)}_{79} Q_5 .
\ee
{}From \eq{q1h}-\eq{q3ph} and \eq{no-d3-nh} we get
\be
\label{18}
Q^{(h)}_{1s} = Q_{1s}
\ee
where $Q_{1s}$ is given by \eq{q1s}. Thus we see that not only do the
parameters $b_{68}$ and $b_{79}$ have the same values here as in the
asymptotically flat case, even the source D1-branes are identical,
despite the total D1-brane charges being very different in the two
cases.

\bigskip

\noindent\underbar{Mass}

The ${1\over 4}$ BPS mass formula in terms of the various charge
densities in this case is
\be
\label{19}
\left({M^{(h)} \over V^{(h)}_{T^4}}\right)^2 = \left({Q^{(h)}_1 \over
V^{(h)}_{T^4}} + Q_5\right)^2 + \left({Q^{(h)}_3\over
\sqrt{g_{77}g_{99}}} - {Q^{(h)'}_3 \over \sqrt{g_{66} g_{88}}}\right)^2
.
\ee
Using \eq{no-d3-nh}-\eq{18} it can be easily seen that
\be
\label{20}
\left(M^{(h)}\right)^2 =  V^{(h)}_{T^4} \left(4Q_{1s}Q_5\right).
\ee
Apart from the extra factor of the $T^4$ volume in the near horizon
geometry, this is exactly the same as \eq{13}. The extra volume factor
correctly takes into account the difference in the 6-dimensional
Newton's constant between the asymptotically flat and near horizon
geometries because of the difference in the $T^4$ volume in the two
cases. We have already seen that the B-field is automatically
self-dual in the near horizon geometry and that the volume of $T^4$
satisfies the condition given by \eq{16} and \eq{17}. We now see that
the mass of the bound state is already at the fixed point value. Thus
the solution we
have here provides an explicit demonstration of the attractor mechanism
of
\cite{Ferrara-et-al}.


\vspace{1em}

\subsection{Motion of a separated D1-brane}

In this subsection we will discuss the fragmentation of the
D1/D5 system in which a single D1-brane splits off. As
mentioned above, the presence of a $B$-field leads to
a bound state. Therefore, if we consider the motion of the
single D1-brane in the background geometry of the
remaining branes, we should find an attractive potential.
This problem was addressed in \cite{SeiWit99a} who
argued that the potential is of the Liouville form
in  the near-horizon geometry. We will
use our supergravity solution to find the attractive
potential in the full geometry which will include
the above result in the near-horizon limit.

Let us assume that the remaining D1/D5 system is given by
the parameters $Q_1,Q_5$. The motion of the
separated D1-brane can then be described by the following
DBI action, coupled to the supergravity solution
\eq{full-solution}-\eq{full-solution-5}
\be
S_1= -{1\over 2\pi g\alpha'} \int d^2\sigma \left(e^{-\phi}
\sqrt{ - {\rm det}(\hat g + 2\pi\alpha' \hat B)}
- C^{(2)} \right)
\ee
where
\be
\hat g_{\alpha\beta} = g_{\mu\nu}\partial_\alpha x^\mu
\partial_\beta x^\nu, \quad
\hat B_{\alpha\beta} = B_{\mu\nu}\partial_\alpha x^\mu
\partial_\beta x^\nu
\ee
represent the pull-backs of the metric and the $B$-field
onto the D1-brane world-sheet. We will consider
a radially moving D1-brane, given by
\be
x^0 =\sigma_0 \equiv \tau, x^5 = \sigma_1
\equiv \sigma, r = r(\tau)
\ee
It is easy to find $S_1$ explicitly for  small $B$-moduli
\footnote{For radial motion, $\hat B$ vanishes, but the
effect of non-zero $B$ shows up in the presence of
non-zero angles $\phi,\psi$ which leads to non-zero $C^{(2)}$.},
(small angles $\phi,\psi$):
\be
S_1 =  {1\over  g\alpha'} \int d\tau \left( \phi\psi f_5^{-1}
\right) + o(v^2)
\ee
As we have argued above, the minimum-energy bound state
requires the condition \eq{12}, which in turn means
$\phi=-\psi$. For small $B$-moduli, therefore, the
coefficient of $f_5^{-1}$ is $-\phi^2 = -b^2$, where
$b= b_{68}= - b_{79}$.

Thus the separated D1-brane feels an attractive potential
\be
V(r) ={1\over  g\alpha'} b^2 (1 + {\mu_5\alpha'\over 2 r^2})^{-1}
\ee
The near-horizon limit is given by the functional
form $r^2$ which, in terms of the scalar field $\phi=
\ln r$ on the throat, is of the Liouville form $e^{2\phi}$.

It is remarkable that the separated D1-brane feels a static potential
which is the hall-mark of a non-BPS situation, whereas normally one
expects a Dp-Dq system with $p-q$ mod $4=0$ to be BPS. It appears even
more remarkable if one recalls that the supergravity solution with the
B-field, which shows the non-zero binding energy, is obtained simply
by a series of T-duality transformations. Appearance of
non-zero binding energy as a result of T-duality has been
noted before (see, e.g., \cite{Breckenridge-Myers-}). See also
related remarks in \cite{SeiWit99b}.

\section{Gauge Theory Description of the D1-D5 Bound State}

In this section we discuss the microscopic theory of the $D1-D5$
system. The moduli space of instantons of the 5-brane gauge theory
with gauge group $U(Q_5)$ and instanton number $Q_1$
is the resolved Sym$_{Q_1Q_5}(\tilde T^4)$.
\footnote{We have
here dropped the subscript `s' on the source D1-brane charge (which
corresponds now to the instanton number) and, furthermore, by abuse of
notation, $Q_5$ and $Q_1$ are now representing the corresponding
numbers of branes rather than appropriate charge densities as in the
previous section.}.
In the limit of small instanton size one can
approximate this description by directly constructing
the gauge theory corresponding to the system of
$Q_1$ D1-branes and $Q_5$ D5-branes \cite{CalMal}, \cite{HasWad97}.
This description provides a good physical picture of the dynamics of the

brane system. The brane configuration is as follows.
The D1-branes are wrapped on the circle $S^1$ along the direction
$x_5$ and the D5-branes are wrapped on the 5-torus $S^1 \times
T^4$. The coordinates of $T^4$ are denoted by $x_i , \ i = 6, 7, 8,
9$. The radius $R$ of $S^1$ is chosen so that $R \gg \sqrt{\alpha'}$
and $V_4 \sim \alpha^{\prime 2}$.

The above configuration of branes interact via open strings that are
attached to the branes. Denote these by $(1,1), \ (5,5) , \ (1,5)$ and
$(5,1)$. The low energy dynamics of this system is described by a
$U(Q_1) \times U(Q_5)$, $N=4$ gauge theory in the 2 dimensions common
to the 2 branes i.e. $(t, x_5)$. We also allow for a non-zero $\theta$
angle corresponding to the relative $U(1)$ that acts on the
hypermultiplets.

The field content is summarized as follows:

\begin{enumerate}
\item[{(i)}] \underbar{$(1,1)$ Strings} :

The fields corresponding to the massless excitations of the $(1,1)$
strings are obtained from the dimensional reduction of $U(Q_1)$, $N=1$
SYM in
10-dim. The bosonic fields can be organized as vector and hyper
multiplets of $N=2$ theory in 4-dim. in the adjoint representation. of
$U(Q_1)$:
\beq
{\rm vector~multiplets} &\colon& A^{(1)}_0 , \ A^{(1)}_5 , \ Y_m \ \ \
\ m = 1, 2, 3, 4 \nonumber \\ [2mm]
{\rm hyper~multiplets} &\colon& Y_i , \ \ \ \ i = 6, 7, 8, 9 . \nonumber
\eeq
In the above description we have neglected the winding modes of the
$(1,1)$ strings on the torus $T^4$.

\item[{(ii)}] \underbar{$(5,5)$ Strings} :

The massless modes of the $(5,5)$ open strings are obtained in a similar
way except that the gauge group is $U(Q_5)$ and we neglect the $KK$
modes corresponding to the torus $T^4$.
\beq
{\rm vector~multiplets} &\colon& A^{(5)}_0 , \ A^{(5)}_5 , \ X_m \ \ \
\ m = 1, 2, 3, 4 \nonumber \\ [2mm]
{\rm hyper~multiplets} &\colon& X_i , \ \ \ \ i = 6, 7, 8, 9 . \nonumber
\eeq

\item[{(iii)}] \underbar{$(1,5)$ and $(5,1)$ Strings} :

The fields corresponding to the massless excitations of the $(1,5)$
and $(5,1)$ open strings can be organized as doublets of the $SU(2)_R$
symmetry of the gauge theory. $SU(2)_R$ is the diagonal subgroup of
$SO(4)_I \simeq SU(2) \times SU(2)$ which acts on the coordinates
$x_i = 6,7,8,9$.
\be
\label{3.1}
{\rm hyper~multiplets} : \chi_{a'a} = \pmatrix{\chi^1_{a'a} \cr
\chi^2_{a'a}} = \pmatrix{A_{a'a} \cr B_{a'a}}
\ee
$A_{a'b}$ and $B_{a'b}$ transform in the bi-fundamental
representation of $U(Q_1) \times \overline{U(Q_5)}$. The $U(1) \times
\overline{U(1)}$ subgroup is important. One combination leaves the
hypermultiplet invariant. The other combination is active and
$(A_{a'a}, B_{a'a})$ have $U(1)$ charges $(+1, -1)$. $\chi$ is a
chiral spinor of $SO(4)_I$ with convention $\Gamma_{6789} \ \chi = -
\chi$.

\end{enumerate}

In the above discussion, the fields $Y_{i}$ and $X_{i}$ along the
torus directions are assumed to be compact. However it is not obvious
how to compactify the range of $\chi$ so that the integration over
this field in the path integral is finite.

\vspace{1em}

\subsection{The Potential Terms}

The potential energy density of the vector and hyper multiplets is a
sum of 4 positive terms,
\beq
\label{3.2}
V &=& V_1 + V_2 + V_3 + V_4 \\ [2mm]
\label{3.3}
V_1 &=& - {1\over 4g^2_1} \sum_{m,n} tr_{U(Q_1)} [Y_m, Y_n]^2 - {1\over
4g^2_5} \sum_{m,n} tr_{U(Q_5)} [X_m, X_n]^2 \\ [2mm]
\label{3.4}
V_2 &=& - {1\over 2g^2_1} \sum_{i,m} tr_{U(Q_1)} [Y_i, Y_m]^2 - {1\over
2g^2_5} \sum_{i,m} [X_i, X_m]^2 \\ [2mm]
\label{3.5}
V_3 &=& {1\over 4} \sum_m tr_{U(Q_1)} (\chi X_m - Y_m \chi) (X_m
\chi^\dagger -
\chi^\dagger Y_m)^2 \\ [2mm]
V_4 &=& {1\over 4} tr_{U(Q_1)} (\chi i \Gamma^T_{ij} \chi^+ + i [Y_i,
Y_j]^+ - \zeta^+_{ij} {1\!\!\!1 \over Q_1})^2 \nonumber \\ [2mm]
\label{3.6}
&& + {1\over 4} tr_{U(Q_5)} (\chi^+ i \Gamma_{ij} \chi + i [X_i,
X_j]^+ - \zeta^+_{ij} {1\!\!\!1 \over Q_5})^2
\eeq
The potential energy $V_4$ comes from a combination of $F$ and $D$
terms of the higher dim. gauge theory. $\Gamma_{ij} = {i\over 2}
[\Gamma_i, \Gamma_j]$ are spinor rotation matrices. The notation
$a^+_{ij}$ denotes the self-dual part of the anti-symmetric tensor
$a_{ij}$.

In $V_4$ we have included the Fayet-Iliopoulos (FI) terms
$\zeta^+_{ij}$, which form a triplet under $SU(2)_R$. Their inclusion
is consistent with $N=4$ SUSY. The FI terms can be identified with the
self dual part of $B_{ij}$, the anti-symmetry tensor of the NS sector
of the closed string theory. This identification at this stage rests
on the fact that (i) $\zeta^+_{ij}$ and $B^+_{ij}$ have identical
transformation properties under $SU(4)_I$ and (ii) at the origin of
the Higgs branch where $\chi = X = Y = 0$, $V_4 \sim \zeta^+_{ij}
\zeta^+_{ij}$. This signals a tachyonic mode from the view point of
string perturbation theory. The tachyon mass is easily computed and
this implies the relation $\zeta^+_{ij} \zeta^+_{ij} \sim B^+_{ij}
B^+_{ij}$.

\subsection{D-Flatness Equations and the Moduli Space}

The supersymmetric ground state (semi-classical) is characterized by
the 2-sets of D-flatness equations which are obtained by setting $V_4
= 0$. They are best written in terms of the $SU(2)_R$ doublet fields
$N^{(1)}_{a'b'}$ and $N^{(5)}_{ab}$ :
\beq
N^{(1)} &=& \pmatrix{N^{(1)}_1 \cr N^{(1)}_2} = \pmatrix{Y_9 + i Y_8
\cr Y_7 + i Y_6} \nonumber \\ [2mm]
\label{3.7}
N^{(5)} &=& \pmatrix{N^{(5)}_1 \cr N^{(5)}_2} = \pmatrix{X_9 + iX_8
\cr X_7 + iX_9}
\eeq

We also define $\zeta=\zeta^{+}_{69}$ and $\zeta_{c}=
\zeta^{+}_{67}+ i\zeta^{+}_{68}$.
With these definitions the 2 sets of D-flatness conditions become:
\subequations
\label{3.8}
\be
(AA^+ - B^+B)_{a'b'} + [N^{(1)}_1, N^{(1)\dagger}_1]_{a'b'} -
[N^{(1)}_2,
N^{(1)\dagger}_2]_{a'b'} = {\zeta \over Q_1} \delta_{a'b'}
\ee
\be
(AB)_{a'b'} + [N^{(1)}_1, N^{(1)\dagger}_2]_{a'b'} = {\zeta_{c} \over
Q_1} \delta_{a'b'}
\ee
\be
(A^+A - BB^+)_{ab} + [N^{(5)}_1, N^{(5)\dagger}_1]_{ab} - [N^{(5)}_2,
N^{(5)\dagger}_2]_{ab} = {\zeta \over Q_5} \delta_{ab}
\ee
\be
(A^+B^+)_{ab} + [N^{(5)}_1, N^{(5)\dagger}_2]_{ab} = {\zeta_{c} \over
Q_5} \delta_{ab}
\ee
\endsubequations

The hypermultiplet moduli space is a solution of the above equations
modulo the gauge group $U(Q_1) \times U(Q_5)$. A detailed discussion
of the procedure was given in \cite{HasWad97}. Here we summarize.

If we take the trace parts of Eqns. \eq{3.8}
we get the {\em same} set of 3
equations as the D-flatness equations for  a $U(1)$ theory
with $Q_1 Q_5$ hypermultiplets, with $U(1)$ charge assignment $(+1,
-1)$ for $(A_{a'b}, B^T_{a'b})$. Thus,
\subequations
\label{3.9}
\be
\sum_{a'b} (A_{a'b} A^\ast_{a'b} - B^T_{a'b} B^{T\ast}_{a'b}) = \zeta
\ee
\be
\sum_{a'b} A_{a'b} B^T_{a'b} = \zeta_c
\ee
\endsubequations
For a given point on the surface defined by Eqns. \eq{3.9}
the traceless parts of \eq{3.8} lead to $3Q^2_1 + 3Q^2_5 - 6$
constraints on
among $4Q^2_1 + 4Q^2_5 - 8$ degrees of freedom corresponding to the
traceless parts of the adjoint hypermultiplets $N^{(1)}$ and
$N^{(5)}$. Using $Q^2_1 + Q^2_5 - 2$ gauge conditions corresponding to
$SU(Q_1) \times SU(Q_5)$ we have $(3Q^2_1 + 3Q^2_5 - 6) + (Q^2_1 +
Q^2_5 - 2) = 4Q^2_1 + 4Q^2_5 - 8$ conditions for the $(4Q^2_1 + 4Q^2_5
-8)$ degrees of freedom in the traceless parts of $N^{(1)}$ and
$N^{(5)}$. The 8 degrees of freedom corresponding to $trX_i$ and
$trY_i$, $i = 6,7,8,9$ correspond for the centre-of-mass
of the D5 and D1 branes
respectively.

The Eqns.\eq{3.9} for $Q_1Q_5$ hypermultiplets have been
recently discussed in the context of instantons in non-commutative
gauge theory \cite{NekSch,SeiWit99b}.
They define $M = T^\ast CP^{Q_1Q_5-1}$, the cotangent
bundle of the complex hyper-K\"ahler manifold $CP^{Q_1Q_5-1}$. The FI
parameters $\zeta^+_{ij}$ can be identified with the moduli of the
hyper-K\"ahler metrics on $M$. For example, in the case of $Q_1Q_5 =2$,
$M$ is
the Eguchi-Hanson space. When the FI parameters go to zero, it reduces
to the
singular space, ${C^2/Z_2}$. Similarly, we can show that when the FI
terms go to zero, $M$ defined by the equations (3.12) and (3.13) becomes
the
singular space (appendix A)
\be
\label{3.10}
M_0 =
C^2 \times {C^2 \over w} \times {C^2 \over w^2} \times \cdots {C^2 \over
w^{N-1}} .
\ee
The singularity of $M_0$ corresponds to the cycle of length $N$ of the
permutation group $S_N$. Associated with this singularity is a chiral
primary
operator with dimension $(h,\bar{h})=(\frac{N-1}{2},\frac{N-1}{2})$.This
result
has appeared before in \cite{SeiWit99a} using a different approach. In
order
to recover the singularity structure corresponding to all the other
cycles and
the corresponding chiral primaries, we may have to consider D-terms
(similar
to equations (3.12) (3.13) with FI parameters equal to zero )
corresponding to
effective $U(1)$ theories describing the splitting processes
\cite{Justin-thesis}
$(Q_1,Q_5) \rightarrow (Q_1',Q_5')+(Q_1'',Q_5'')$.

We would like to mention that all the chiral primaries have been
obtained
using the symmetric product description of the instanton moduli space on
$T^4$.
\cite{DavManWad99}

\vspace{1em}

\subsection{The Bound State in the Higgs Phase}

Having discussed the moduli space that characterizes the SUSY ground
state we can discuss the fluctuations of the transverse vector
multiplet scalars $X_m$ and $Y_m$, $m = 1,2,3,4$. In the Higgs phase
since $\langle X_m \rangle = \langle Y_m \rangle = 0$ and $\chi =
\overline \chi$ lies on the surface defined by Eqn. \eq{3.9}.
The relevant
action of fluctuations in the path integral is,
\be
S = \sum_m \int dt dx_5 (tr_{U(Q_5)} \partial_\alpha X_m
\partial^\alpha X_m + tr_{U(Q_1)} \partial_\alpha Y_m \partial^\alpha
Y_m) + \int dt dx_5 (V_2 + V_3)
\ee

We restrict the discussion to the case when $Q_5=1$ and $Q_1$ is
arbitrary.
In this case the matrix $X_m$ is a real number which we denote by $x_m$.

$\chi$ is a complex column vector with components $(A_{a'},B_{a'})$,
$a'= 1,..
.,Q_1$. Since we are looking at the fluctuations of the $Y_m$ only to
quadratic
order in the path integral, the integrals over the different $Y_m$
decouple
from each other and we can treat each of them separately. Let us discuss
the
fluctuation $Y_1$ and set $(Y_1)_{a'b'}=\delta_{a'b'} y_{1{a'}}$. Then
the potential
$V_3$, \eq{3.5} becomes
\beq
V_3= \sum_{a'} (|A_{a'}|^2 + |B_{a'}|^2 )(y_{1a'} -x_1)^2
\eeq
We will prove that $|A_{a'}|^2 + |B_{a'}|^2$ can never vanish if the FI
terms
are non-zero.
In order to do this let us analyze the complex D-term equation \eq{3.9}
\beq
A_{a'}B_{b'}+ [N^{(1)}_1,N^{(1)\dagger}_2]_{a'b'} = {\zeta_c \over Q_1}
\delta_{a'b'}
\eeq
We can use the complex gauge group $GL(C,Q_1)$ to diagonalize the
complex
matrix $N^{(1)}_1$ \cite{Witten}. Then, the above equation becomes
\beq
A_{a'}B_{b'}+ (n_{a'} - n_{b'})(N^{(1)\dagger}_2)_{a'b'} ={\zeta_c \over
Q_1}
\delta_{a'b'}
\eeq
For $a' \neq b'$, this determines the non-diagonal components of
$N^{(1)}_2$
\beq
(N^{(1)\dagger}_2)_{a'b'} = - { A_{a'}B_{b'} \over n_{a'} - n_{b'} }
\eeq
For $a=b$, we get the equations
\beq
A_{a'}B_{a'}={\zeta_c \over Q_1} , a'= 1,..,Q_1
\eeq
which imply that
\beq
|A_{a'}| |B_{a'}| = { |\zeta_c| \over Q_1}
\eeq
with the consequence that $|A_{a'}|$ and  $|B_{a'}|$ are non-zero for
all
$a' = 1,..,Q_1$. This implies that $(|A_{a'}|^2 + |B_{a'}|^2) > 0)$, and
hence
the fluctuation $(y_{1a'} - x_1)$ is massive. If we change variables
$y_{1a'}
\rightarrow y_{1a'} +x_1$, then $x_1$ is the only flat direction. This
corresponds to the global translation of the 5-brane in the $x_1$
direction.

A similar analysis can be done for all the remaining directions
$m=2,3,4$ with
identical conclusions. This shows that a non-zero FI term implies a true
bound
state of the $Q_5=1$, $Q_1=N$ system.
This result was previously presented in \cite{Spain-talk}.
If $FI=0$, then there is no such guarantee and the
system can easily fragment, due to the presence of flat directions in
$(Y_m)_{a'b'}$.

What the above result says is that when the FI parameters are non-zero
the zero mode of the fields $(Y_m)_{a'b'}$
is massive. If we regard the zero mode as a collective coordinate then
the Hamiltonian of the zero mode has a quadratic potential which agrees
with
the near horizon limit of the Liouville potential derived in section
2.3.

The general case with an arbitrary number of $Q_1$ and $Q_5$ branes
seems significantly harder to prove but we are optimistic.

\vspace{1em}
\section{SCFT on the Moduli Space of Non-Commutative Gauge Theory}

Here we briefly explore the possible connection of this work with
Non-commutative geometry. As has been discussed by several authors,
the presence of a non-zero $B_{NS}$ in the near horizon geometry of a
D-brane system wrapped on a torus, will modify the boundary conditions
of the open strings along the brane directions. Then, the brane world
volume Yang-Mills theory is substituted by a Yang-Mills theory on the
non-commutative torus.

In the D1/D5 system, the configurations which break one half of the
supersymmetries of the 5-brane theory are self-dual connections of this
NCYM theory. The moduli space of these connections with the $B_{NS}$
turned on,
will then define a conformally invariant sigma-model.

The $N=4$ symmetric product SCFT with $c = 6Q_1Q_5$ has been constructed
in
some detail and its $20$ dim. moduli space can be explicitly
constructed in terms of operators of the SCFT \cite{DavManWad99}.
Of the four operators
corresponding to the blowup modes of the orbifold singularity, $3$
correspond
to $B^{(+)}_{ij}$, the self dual part of the $B_{ij}$ in the NS sector
of the
closed string theory and the remaining one corresponds to the $\theta$
parameter.

We wish to propose that this new sigma-model, coming from the
noncommutative
Yang-Mills theory, is at some point in the moduli space of
the symmetric product SCFT, deformed by the addition of the four
marginal
deformations corresponding to the moduli $B^{+}_{ij}$ and $ aC_0 +
bC_4$.

After this work was completed we became aware of \cite{Mikh} which has
some overlap with this work.

\vspace{5ex}

{\center {\large \bf Acknowledgement}}

We would like to acknowledge Justin David for collaboration during the
early stages of this work, and his critical comments on the
manuscript.  We would like to thank Fawad Hassan for a very helpful
correspondence regarding \cite{Fawad}.
We (SRW,KPY) would also like to thank the
'Extended Workshop in String Theory', at the Mehta Research Institute,
for hospitality during the final stages of this work.
We thank Ashoke Sen
and David Tong for useful comments on some aspects of this work
related to the moduli space of instantons.

\appendix

\section{Derivation of \eq{3.10}}
We now study the limiting case of $M = T^\ast CP^{Q_1Q_5-1}$ as
$\zeta \rightarrow 0$. However this seems presently hard to do and hence
we set
$\zeta = 0$ in Eqn. \eq{3.9}, and explicitly demonstrate
that these equations describe the symmetric product of $(Q_1 Q_5 -1)$
copies of $C^2 = C \times C$ where $C$ denotes the complex plane.

In the following we will, for convenience,
use the symbol $N$ for $Q_1 Q_5$.

Since we are dealing with $N \ U(1)$ hyper multiplets, it is
convenient to denote the hypermultiplets by $A_a,
B^+_a , \ a = 1, \cdots , N$. Then Eqns.
\eq{3.9} become (we have set
$\zeta = 0)$,
\subequations
\label{3.17}
\be
\label{3.17a}
\sum^{N}_{a=1} (|A_a|^2 - |B_a|^2) = 0
\ee
\be
\label{3.17b}
\sum^{N}_{a=1} A_a B_a = 0
\ee
\endsubequations
The hypermultiplets constitute  $2N$ complex, or equivalently
$4N$ real variables. These variables  are not invariant
under the $U(1)$ ($A$ has $U(1)$ charge 1 and $B$ has
charge $-1$).
Using \eq{3.17a}  and dividing by $U(1)$ leaves
us with $4N-2$ real, or $2N-1$ complex variables which can be
parametrized by the $2(N-1)+N=3N-2$ (complex) gauge invariant variables
\beq
P_a &=& A_aB_a , \ \ \ \ a = 1, \cdots , N \nonumber \\ [2mm]
M_a &=& A_aB_N , \ \ \ \ a = 1, \cdots , N-1 \\
N_a &=& B_aA_N , \ \ \ \ a = 1, \cdots , N-1 \nonumber
\eeq
subject to the $N-1$ obvious conditions
\[
P_a P_N = M_aN_a , \ \ \ a = 1, \cdots , N-1
\]
Let us now use the complex equation \eq{3.17b}, which translates
to
\[
\sum_a^{N} P_a=0
\]
The $N$ constraints involving the $3N-2$ variables $P,M,N$
can be solved in terms of $(N-1)$ {\em pairs} of complex numbers
$(\eta_i , \xi_i), i = 1, \cdots , (N-1)$
\beq
P_a &=& w^a P_N , \ \ \ \ w^N = 1,\ \ P_N = \prod^{N-1}_{i=1}
(\eta_i \xi_i)^{1/2} \nonumber \\ [2mm]
M_a &=& \sqrt{w^a_N} \eta_a \prod^{N-1}_{j\neq a} (\eta_j
\xi_j)^{1/2} \\ [2mm]
 N_a &=& \sqrt{w^a_N} \xi_a \prod^{N-1}_{j\neq a} (\eta_j
\xi_j)^{1/2} \nonumber
\eeq
This solution is characterized by a point in $(N-1)$ copies of
$C^2$. However there are identifications which can be
characterized in terms of the cycles of the symmetric group
$S_N$. For example consider the action of the largest cycle
corresponding to the diagonal action $(\xi_a, \eta_a) \rightarrow
e^{i 2\pi a/N} (\xi_a, \eta_a) \equiv
w^a (\xi_a, \eta_a)$ where $w^N=1$.

It is clear that such a transformation leaves the gauge invariant
coordinates $P_a$, $M_a$ and $N_a$ invariant. Hence the $(N-1)$ copies
of $C^2$ have to be quotiented as follows:
\beq
C^2 \times {C^2 \over w} \times {C^2 \over w^2}
\times \cdots {C^2 \over w^{N-1}}
\eeq

\end{document}